\documentclass{aa}  
\usepackage{graphics} 
\usepackage{rotating}
\usepackage{natbib}
\bibpunct{(}{)}{;}{a}{}{,} % to follow the A&A style
\def\kms{~km~s$^{-1}$}
\def\cmmt{~cm$^{-3}$}
\def\cmmd{~cm$^{-2}$}

\def\mum{$\mu$m}
\def\ui{ergs\,s$^{-1}$\,cm$^{-2}$\,sr$^{-1}$}
\newcommand{\ratio} {N({\rm H}_2) / I_{\rm CO}}

\def\HII{H{\sc ii}}
\def\HI{H{\sc i}}

\def\OI{[O{\sc i}]}
\def\CII{[C{\sc ii}]}
\def\Ha{H$\alpha$}
\def\ga{\lower.5ex\hbox{$\; \buildrel > \over \sim \;$}}
\def\la{\lower.5ex\hbox{$\; \buildrel < \over \sim \;$}}

\def\e{$\pm$}
\begin{document}
   \title{[CII]  emission and star formation in  the spiral arms of M31
   \thanks{Based on observations with ISO, an ESA project with instruments 
   funded by ESA Member States (especially the PI countries: France, Germany, 
   The Netherlands and the UK) and with the participation of ISAS and NASA.   }} 

%   \subtitle{I. Overviewing the $\kappa$-mechanism}

   \author{N. J. Rodriguez-Fernandez
          \inst{1, 2}
          \and
         J. Braine\inst{1}
         \and
          N. Brouillet \inst{1}
          \and
          F. Combes \inst{3}
          }

   \offprints{N.J. Rodriguez-Fernandez}

   \institute{Observatoire de Bordeaux, L3AB (UMR 5804),  CNRS/Universit\'e Bordeaux 1,  BP 89, 2 rue de l'Observatoire, 33270 Floirac, France \\ 
     \email{nemesio.rodriguez@obs.u-bordeaux1.fr}
     \and Universit\'e Denis Diderot (Paris VII) \& Observatoire de Paris,  61 Av de  l'Observatoire, F-75014 Paris, France
            \and
             LERMA, Observatoire de Paris, 61, Av. de l'Observatoire, 75014, Paris, France\\ }

   \date{Received Dec. 8, 2005; accepted }
\authorrunning{Rodriguez-Fernandez et al.}
\titlerunning{[CII] emission in M~31}
% \abstract{}{}{}{}{} 
% 5 {} token are mandatory
%====================================

 \abstract
 {The \CII\ 158 \mum\ line is the most important coolant of the interstellar medium 
in galaxies but substantial variations are seen from object to object. 
The main source of the emission at a galactic scale is still poorly understood and candidates range from photodissociation regions (PDRs) to the cold neutral or diffuse warm ionized medium.  
Previous studies of the \CII\ emission in galaxies have a resolution of several kpc or more so the observed emission is an average of different ISM components.}
{The aim of this work is to  study, for  the first time, the \CII\ emission at
the scale of a spiral arm. We want to investigate the origin of this
line and its use as a tracer of star formation.}
{ We present \CII\ and \OI\  observations of a 
%$5' \times 20'$ 
segment of a spiral arm of M~31 using the Infrared Space Observatory.  The \CII\ 
emission is compared with tracers of neutral gas (CO, \HI) and star formation 
(H$\alpha$, Spitzer 24 \mum).}
{
%The \CII\ emission is clearly much more linked to the level of star formation, whereas the correlation with the neutral gas is much weaker.  
The similarity of the \CII\ emission with the \Ha\ and 24 \mum\ images 
is striking when smoothed to the same resolution, whereas the correlation with the neutral gas is much weaker. 
%In fact, the \CII\ emission per  H-atom increases dramatically in zones of active star formation within the spiral arm.  
The \CII\ cooling rate per H atom increases dramatically from  $\sim 2.7 \,10^{-26}$ ergs\,s$^{-1}$\,atom$^{-1}$ in the
border of the map to $\sim 1.4 \,10^{-25}$ ergs\,s$^{-1}$\,atom$^{-1}$ in the regions of star formation. 
The \CII/FIR$_{42-122}$  ratio is almost constant at 2\%, a factor 3 higher than typically quoted. 
However, we do not believe that M~31 is unusual.  Rather, the whole-galaxy  fluxes used for the comparisons include the central regions where the \CII/FIR  ratio is known to be lower and the resolved observations neither isolate a  spiral arm nor include data as far out in the galactic disk as the observations  presented here.
%Examining older data shows that the tendency for higher \CII/FIR ratios in the outer disk is also present. 
A fit to published PDR models yields a plausible average solution of $G_0 \sim 100$  and $n \sim 3000$ for the PDR emission in the regions of star formation in the  arm of M31.}
{} 
\keywords{Galaxies: spiral -- Galaxies: ISM -- Infrared: ISM -- stars: formation -- ISM: molecules -- Galaxies: individual (Messier 31)}

\maketitle

%========================================
 
%________________________________________________________________

\section{Introduction}
The  \CII~ 158 \mum\ fine structure line is the strongest spectral line in the universe,
carrying typically close to 1\% of the energy emitted in the  Far-IR from galaxies.
%(REF -- Loeb?).
This line has a critical density for excitation around 1000 cm$^{-3}$,
and is therefore widespread in galaxies.  Previous studies of \CII\ emission in 
spiral galaxies include \citet{Madden93}, \citet{Nikola01}, \citet{Kramer05}, 
and \citet{Braine_n4414c} for
respectively NGC 6946, M 51, M~83, and NGC 4414, and \citet{Malhotra01} for a statistical
study.  All of these have a resolution of several kpc and thus average
a wide variety of environments within the beam.  
Carbon is ionized more easily than Hydrogen so \CII\ emission, although from an ion, 
can arise from ionized \HII~ regions to
mainly neutral photo-dissociation regions \citep[PDR; e.g. ][]{Tielens85a}  at 
the border of molecular clouds, exposed to UV radiation from young massive stars.
A significant contribution could also come from the diffuse atomic 
interstellar gas, as shown by \citet{Crawford85} and \citet{Madden93},
or the diffuse ionized medium \citep{Heiles94}.  
However, the main contributions are not known in detail, especially
at high spatial resolution. 
%It is interesting to note the highly
%varying ratio between the CO(1--0) and CII intensities, varying from 
%about 1300 for galactic disks \citep{Nakagawa98,Braine_n4414c}, to 
%about 6000 in starbursts \citep{Stacey91} and up to 23000
%for the Large Magellanic Cloud \citep{Mochizuki94}.
%Even in the same galaxy, this ratio can vary in large proportions.

Far from the first study of \CII~ emission, the present work is unique in
that we study emission at the scale of a spiral arm, only possible in the very
nearby spirals M~31 or M~33.
M~31 is the nearest density-wave spiral galaxy, where we can test
large-scale dynamics and star formation.  It is the only density-wave galaxy 
where spiral arms are resolved by ISO (M33 being more stochastic), as shown in Fig. 1. 
The scale of the \CII\ observations presented here is new and allows us to 
study the spiral arm environment separate from the rest of the galaxy.
With a focus on testing the origin of the \CII\ line, 
we compare the \CII\ emission across an arm in M 31 with the CO and HI, tracing the
neutral gas, the H$\alpha$ emission tracing ongoing star formation but suffering 
from extinction, and the Spitzer 24\mum\ image tracing the warm dust
heated by recent star formation.  
\CII\ emission as a tracer of star formation has 
the advantage of not being affected by interstellar extinction. 

\section{Observational data}

We have observed a portion of the northern spiral arm of M31 in the \CII \ 157.7 \mum~
and \OI \ 63.8 \mum\ lines using the {\it Long Wavelength Spectrometer} (LWS) onboard
the {\it Infrared Space Observatory } (ISO).  The observed region, which is shown in
Fig. 1 on a 175 \mum\ image by \citet{Haas98}, is the most intense region of star 
formation in M 31
and contains a large concentration of \HII\ regions \citep{Devereux94}.  The \CII \
line was observed with two raster maps of $10 \times 4$ points centered at 
RA=$00h44m38.736s$ and Dec=$41^\circ27'17.208''$ (J2000) and $12 \times 4$ points
centered at RA=$00h45m7.514s$ and Dec=$41^\circ 35' 46.896''$ (J2000).
 The ISO observation numbers 
 (TDT numbers) of those observations are 58001701 and 58001801, respectively. 
 In addition, we have mapped the \OI  \ 63.8 \mum\ line in the central positions of the \CII\
maps with  two small rasters of $2 \times 2$ points  (TDTs 58001703 and 58001903). The
raster maps were  oriented along the spiral arm at an angle of 63 deg. For all the
maps, the observed points are separated by 1'.  
Tables \ref{tab_cii_1} and \ref{tab_cii_2} give the exact observed positions.

The lines were scanned in the LWS 02 mode at a spectral resolution of 0.29  and 0.6
\mum \ for the  \OI\ 63 \mum\ and the \CII \ 158 \mum\ lines, respectively.  The
effective aperture of the LWS detectors is $87^{''}$ for the \OI \ line and $69.4^{''}$
for the \CII \ line \citep{Gry03}.  The observations were processed through the
off-line processing (OLP) software version 10. Further reduction was done with ISAP.
The data reduction  consists basically in dropping bad data points, shifting  the
different scans taken with each detector to a common level and averaging all the data
for each detector. In addition, with ISAP we have defringed the spectra to get rid of
the well-known interference pattern for the long wavelength detectors. We fitted order
1 baselines to the spectra in the vicinity of the lines and Gaussian curves to the
lines using ISAP.  
Tables \ref{tab_cii_1} and \ref{tab_cii_2} give  the \CII\ line fluxes and errors as derived
 from the Gaussian fits.
The absolute flux calibration uncertainties for the LWS01 
mode are smaller than 20\%  (Gry et al. 2003).

We also present an archival {\it Spitzer Space Telescope} image of the same region 
taken with the MIPS instrument at 24 \mum, with a spatial resolution of about $5"$.
The other data sets used here are H$\alpha$ from \citet{Devereux94}, HI from 
\citet{Brinks84}, CO(1--0) emission from \citet{Nieten06}, and the combined 60 
and 100 \mum\ IRAS maps.

\begin{figure}[h!]
\begin{center}
\includegraphics[angle=270,width=8cm]{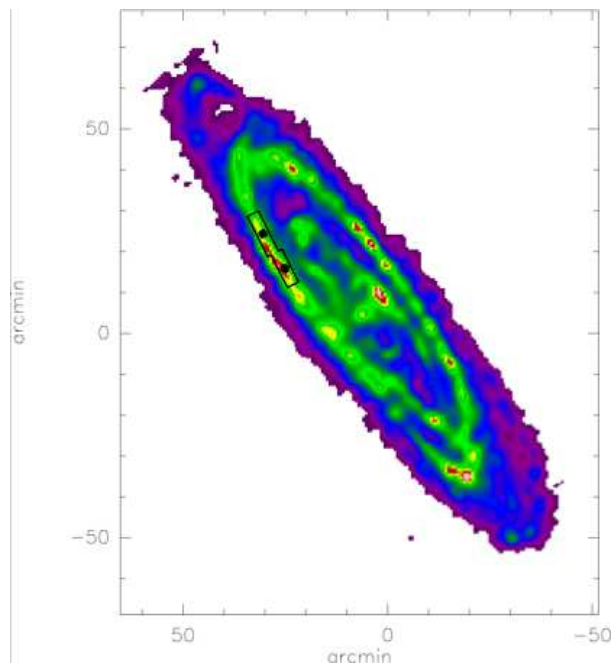}
\caption{Thermal dust continuum emission at 175 \mum\ in M 31 from 
Haas et al. (1998).  The two empty rectangles show the region mapped in the \CII \ 
line while the two solid squares represent the region mapped in the \OI \ line.}
\label{fig1}
\end{center}
\end{figure}

\begin{table*}
\caption{Observed positions in the TDT=58001701 raster map ($\Delta \alpha$ and $\Delta \delta$ offsets in arcmin with respect to  RA=$00h44m38.736s$ and Dec=$41^\circ27'17.208''$, J2000) and \CII \ line flux in units of $10^{-20}$ W\,cm$^{-2}$. }
\label{tab_cii_1}
\begin{tabular}{llllllllllll}
\hline
$\Delta \alpha$ & $\Delta \delta$ & \CII & $ \Delta \alpha$ & $\Delta \delta$ & \CII & $\Delta \alpha$ & $\Delta \delta$ & \CII & $\Delta \alpha$ & $\Delta \delta$ & \CII  \\
\hline
-3.37 & -3.33 &   7.0 \e   0.8  &-2.47 & -3.78 &  10.1 \e   0.7  & -1.59 & -4.23 &  14.5 \e   0.5& -0.69 & -4.69 &   8.7 \e   0.4  \\
-2.92 & -2.45 &   7.4 \e   0.8  &-2.02 & -2.89 &  10.2 \e   0.4  & -1.13 & -3.35 &  16.1 \e   0.1& -0.23 & -3.79 &  10.1 \e   0.5 \\
-2.47 & -1.56 &   6.1 \e   0.7  &-1.57 & -2.00 &  13.0 \e   0.8  & -0.68 & -2.46 &  23.1 \e   0.6&  0.22 & -2.90 &  11.6 \e   0.9 \\
-2.01 & -0.66 &   5.1 \e   0.7  &-1.12 & -1.10 &  12.2 \e   0.3  & -0.23 & -1.57 &  21.0 \e   2.0&  0.67 & -2.00 &  16.5 \e   0.6 \\
-1.56 &  0.23 &   6.3 \e   0.8  &-0.67 & -0.21 &  10.5 \e   0.5  &  0.22 & -0.68 &  18.3 \e   0.7&  1.12 & -1.11 &  20.0 \e   1.0 \\
-1.11 &  1.12 &   5.2 \e   0.8  &-0.22 &  0.68 &   9.5 \e   0.7  &  0.67 &  0.22 &  21.7 \e   0.4&  1.57 & -0.22 &  18.6 \e   1.3 \\
-0.66 &  2.01 &   6.6 \e   0.2  & 0.24 &  1.57 &   7.8 \e   0.3  &  1.12 &  1.11 &  20.0 \e   0.6&  2.02 &  0.67 &  18.2 \e   0.5 \\
-0.21 &  2.91 &   8.9 \e   0.8  & 0.69 &  2.47 &   9.2 \e   0.5  &  1.57 &  2.00 &  18.8 \e   0.6&  2.47 &  1.56 &  16.9 \e   0.9 \\
 0.24 &  3.79 &   4.0 \e   2.0  & 1.14 &  3.36 &   8.5 \e   0.1  &  2.03 &  2.89 &  15.9 \e   0.4&  2.93 &  2.46 &  17.8 \e   0.4 \\
 0.69 &  4.69 &   4.6 \e   0.3  & 1.58 &  4.25 &   8.3 \e   0.7  &  2.48 &  3.79 &  17.7 \e   0.8&  3.37 &  3.34 &  18.3 \e   1.0 \\
\hline
\end{tabular}
\end{table*}

\begin{table*}
\caption{Observed positions in the  TDT=58001801 raster map ($\Delta \alpha$ and $\Delta \delta$ offsets in arcmin with respect to RA=$00h45m7.514s$ and Dec=$41^\circ 35' 46.896''$, J2000) and \CII \ line flux in units of $10^{-20}$ W\,cm$^{-2}$.}
\label{tab_cii_2}
\begin{tabular}{llllllllllll}
\hline
$\Delta \alpha$ & $\Delta \delta$ & \CII & $\Delta \alpha$ & $\Delta \delta$ & \CII & $\Delta \alpha$ & $\Delta \delta$ & \CII & $\Delta \alpha$ & $\Delta \delta$ & \CII  \\
\hline
-3.82 & -4.23 &   8.0 \e   0.7 &-2.92 & -4.67 &  18.1 \e   0.5  & -2.04 & -5.13 &  17.3 \e   1.4 & -1.14 & -5.58 &  13.0 \e   1.4 \\
-3.37 & -3.35 &   6.6 \e   0.4 &-2.47 & -3.78 &  16.8 \e   1.1  & -1.59 & -4.25 &  28.1 \e   0.5 & -0.69 & -4.68 &  16.2 \e   0.8 \\
-2.92 & -2.45 &   5.4 \e   0.2 &-2.02 & -2.89 &  11.8 \e   1.4  & -1.13 & -3.35 &  18.4 \e   0.6 & -0.23 & -3.79 &  12.2 \e   0.5 \\
-2.47 & -1.56 &   4.5 \e   0.4 &-1.57 & -2.00 &   7.2 \e   0.6  & -0.68 & -2.46 &  11.6 \e   0.8 &  0.22 & -2.90 &   8.8 \e   0.7 \\
-2.02 & -0.66 &   4.6 \e   0.4 &-1.12 & -1.10 &   5.9 \e   0.7  & -0.23 & -1.57 &  13.7 \e   1.2 &  0.67 & -2.01 &   8.5 \e   0.9 \\
-1.57 &  0.22 &   6.2 \e   0.6 &-0.66 & -0.21 &   7.6 \e   0.7  &  0.22 & -0.68 &  14.5 \e   0.7 &  1.12 & -1.11 &   9.2 \e   0.8 \\
-1.11 &  1.12 &   6.0 \e   0.7 &-0.22 &  0.68 &  11.9 \e   0.6  &  0.67 &  0.22 &  14.9 \e   0.7 &  1.57 & -0.22 &   9.9 \e   0.5 \\
-0.66 &  2.01 &   8.3 \e   0.4 & 0.24 &  1.57 &  17.9 \e   1.0  &  1.12 &  1.11 &  23.0 \e   0.6 &  2.02 &  0.67 &   8.8 \e   0.7 \\
-0.21 &  2.91 &   8.5 \e   0.2 & 0.69 &  2.47 &  13.7 \e   0.9  &  1.57 &  2.00 &  19.2 \e   0.5 &  2.47 &  1.56 &   7.1 \e   1.0 \\
 0.24 &  3.79 &   8.3 \e   0.8 & 1.14 &  3.36 &  10.7 \e   0.3  &  2.03 &  2.89 &  12.5 \e   0.9 &  2.92 &  2.46 &   5.9 \e   0.6 \\
 0.69 &  4.69 &   6.6 \e   0.3 & 1.59 &  4.25 &  10.1 \e   0.7  &  2.48 &  3.79 &  10.8 \e   0.7 &  3.38 &  3.34 &   7.3 \e   0.8 \\
 1.14 &  5.58 &   4.4 \e   0.6 & 2.03 &  5.14 &   7.8 \e   0.5  &  2.93 &  4.68 &   7.5 \e   1.4 &  3.82 &  4.23 &   6.1 \e   0.7 \\
\hline
\end{tabular}
\end{table*}

\section{Origin of the \CII\ emission}

Figure 2 shows the contour map of the \CII  \ line integrated intensity 
overlaid on different images of other tracers of neutral (\HI, CO) and ionized 
gas (H$\alpha$) and dust continuum emission at 24 \mum\ and the combined 60 
and 100 \mum\ IRAS maps.  All the images have been reprojected using an orthographic
projection with center in RA=$00h44m51.960s$ and Dec=$ 41^\circ 31' 4.44''$ (J2000)
and rotated by 63 deg. This is the coordinate system that we use in the following.
The right panels show the \CII \ map superposed on the same tracers but smoothed to the
same resolution as the \CII \ (about $70''$).
In addition, since one of the goals of
the observations was to study how the \CII\ emission varied across a spiral arm,  we have
also made a set of cuts perpendicular to the arm in all the tracers smoothed to the same
angular resolution (Fig. 3).
All the peaks have been normalized to the maximum intensity of the \CII\ line. 

\begin{figure*}[t!]
\begin{center}
\includegraphics[angle=0,width=17cm]{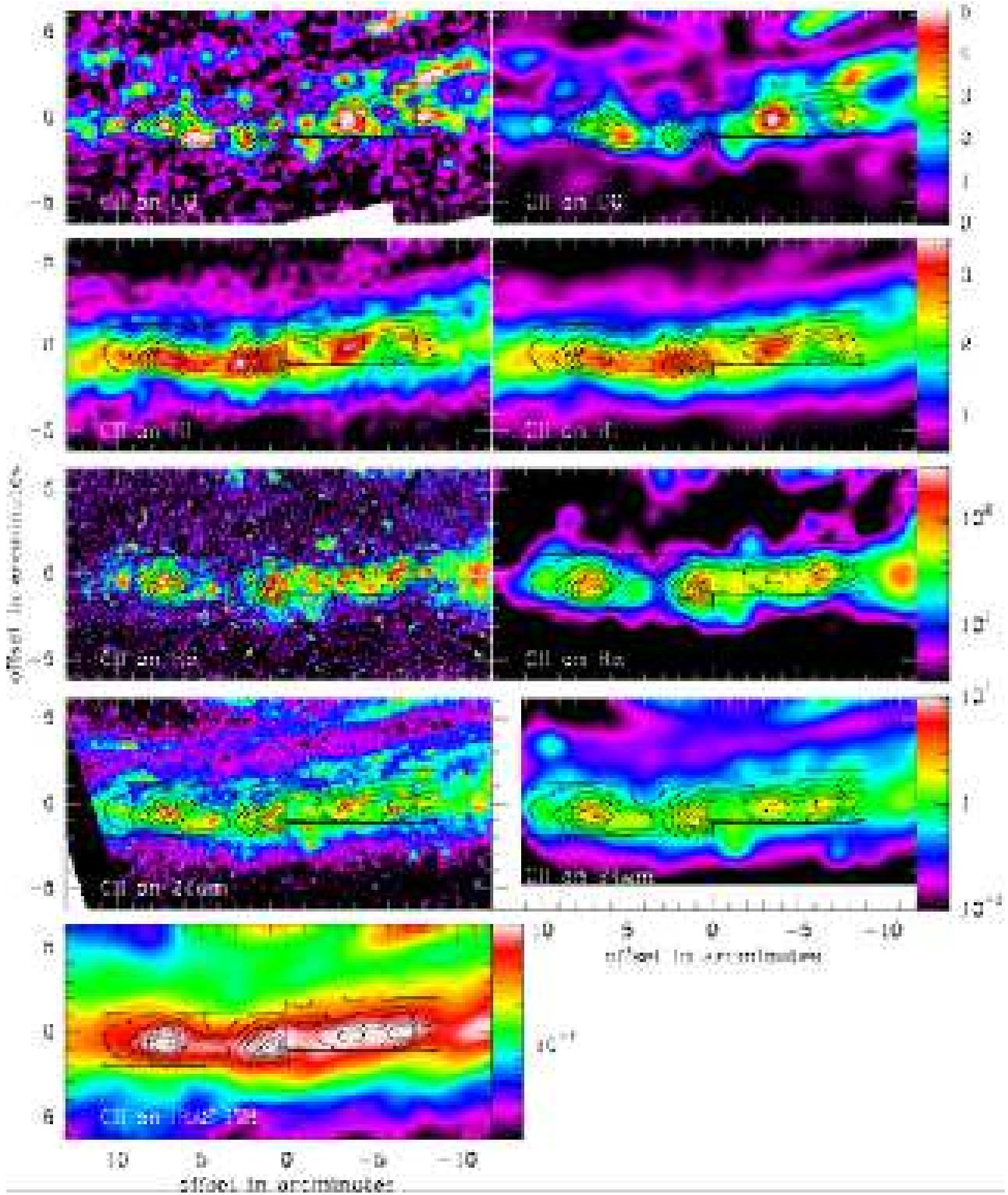}
\caption{CII map (black contours) on the CO(1-0), HI, H$\alpha$, 24 \mum,
and combined $2.58 S_{60} + S_{100}$ IRAS images.
Panels on the left: original resolution. Panels on the right: smoothed to the angular
resolution of the CII observations (70") and with the same transfer function shown
in the wedge to the right.  \CII\ contours are at 3, 6, 9, 12, 15, 18, 21, 24, and
27 $\times 10^{-13}$ ergs cm$^{-2}$ s$^{-1}$.  Units for the maps are in K\kms\ for
the CO(1--0), $10^{21}$ cm$^{-2}$ for the \HI, arbitrary units for the Halpha (log
scale)), MJy/sr for the 24 \mum, and in units of $10^{-3}$ ergs cm$^{-2}$ s$^{-1}$ sr$^{-1}$
for the IRAS map.  The ISM tracers (CO, \HI) are shown with a linear scale and the
tracers of star formation (H$\alpha$, 24 \mum, and IRAS) are shown with a
logarithmic scale.}
\label{fig_maps}
\end{center}
\end{figure*}

Inspection of the maps in Fig. 2 shows that the \CII\ emission peaks are 
typically {\it not} coincident with the CO maxima, although they generally trace the same
part of the spiral arm.  The upper right corner of the \CII\ map shows little emission 
despite a maximum in CO.  Interestingly, the molecular gas in that area appears rather
quiescent as it is generally weak in the tracers of star formation.  
This is clear in Fig. 3.
Single-peaked CO profiles (for example x=4', 0.8', -3.5') are symmetrical with
respect to the arm crest and very similar to the \Ha \ and 24 \mum\ ones, suggesting that
the regions with strong CO emission and little star formation are intrinsically
different (see for instance x=9', y=1.5').  
% After looking at all the double-peaked spectra, there are roba
%The spiral structure of M 31 remains terribly unclear but it is possible that
%these areas (all on the inner side of the arm) may be the zones where the HI is compressd
%to form H$_2$, with the subsequent star formation being visible in the main part of the
%arm.

It is immediately clear in Fig. 2 that the \CII\ emission in the arm follows the tracers of
star formation (H$\alpha$, 24 \mum), rather than the neutral gas (CO, \HI), such that
little of the emission in a spiral arm environment comes from the cold neutral medium
(CNM) as proposed by \citet{Bennett94} and \citet{Wolfire95}.  
This can also be seen in the right panels, where there is excellent agreement between 
the \CII\ contours and the
smoothed H$\alpha$ and 24 \mum\ images.  The solution of the
inverse problem is not unique but the agreement is so good, in particular
with the smoothed 24 \mum, that a high angular resolution \CII\
image should be very similar to the Spitzer image of the warm dust emission.
%[and in our Galaxy...?}

The high angular resolution H$\alpha$ and 24 \mum\  images show many discrete sources 
in addition to an extended component.  Most of the sources seen in the warm dust emission 
at 24 \mum\ are detected in H$\alpha$ as well in spite of the extinction. 
Figure 4 shows the contours of H$\alpha$ flux overlaid on 
the Spitzer 24 \mum\ emission -- the similarity is striking.  This confirms the result
by \citet{Devereux94} -- but at 5$''$ resolution rather than the $\sim 100''$ of IRAS --
that despite the extinction expected for a spiral arm, the 
H$\alpha$ line flux yields a very similar picture of the star formation as far-infrared (FIR) continuum,
showing that even the young regions of star formation are not so dust-enshrouded that
they are not bright in the H$\alpha$ line.
%Finally, one should note that the width at half maximum of the spiral armis $\sim 2'$, which corresponds to a linear distance of 450
%pc, in excellent agreement with the estimate of the \citet{Sauty98} model of the mean
%free path  of the UV photons.

\section{The \CII/FIR ratio}
We have used  the IRAS fluxes to calculate the \CII/FIR ratio flux in the same way as
earlier works. It should be noted that ``FIR" here
refers to the 42 -- 122 \mum\ flux, roughly half the total thermal dust emission.
The \CII/FIR$_{42-122}$ ratio in the mapped region is almost constant at 2\%, well above the value in  the central regions of M~31  \citep[0.6 $\%$;][]{Mochizuki00}.
In the Milky way, the \CII/FIR ratio also increases from 0.2$\%$ in the center \citep[][]{Rodriguez04}
to 0.6$\%$ in the disk   \citep[][]{Nakagawa95}.
Regarding other external galaxies, few data of sufficient sensitivity are
available. In the case of NGC6946 \citep[][]{Madden93}, the \CII/FIR ratio increases from 0.2 $\%$ in the center to 0.9 $\%$ in the northern spiral arm (at a distance of $\sim 4.5$ kpc).
In the observations of M51 reported by \citet{Nikola01},  the \CII/FIR ratio varies from 0.6 $\%$ in the nucleus to 1.3 $\%$ in the spirals arms located at $\sim 5.5$ kpc from the center.
Our data confirm that the \CII/FIR ratio increases with radius in galactic disks and the high \CII/FIR ratio measured here is found at
$\sim 12$ kpc from the center.
The study of the dense molecular gas emission by \citet{Brouillet05} has already shown that the physical properties of the interstellar medium  change with galactocentric distance in M31, as was found by \citet{Sodroski97} for the Milky Way.

%The high \CII/FIR ratio measured in the spiral arm of M31 can reflects that the region presented in this work is at  substantially greater radius than any of the above studies ().

However, it is unclear what physical mechanisms are responsible for this behavior.
Statistical studies of different galaxies \citep{Malhotra01} have shown a tendency for the \CII/FIR ratio to decrease with increasing IRAS 60/100 ratios, i.e. when the dust temperature increases.  
As discussed by \cite{Contursi02}, this effect is also weakly present along the disks of NGC1313 and NGC6946. 
A possible explanation is that  strong UV fields heat the dust to higher temperatures (high 60/100 ratio) than weaker fields but they can also increase the positive charge of the dust grains and 
reduce the efficiency of the photoelectric effect, which is the main heating mechanism in the region of the PDRs where the \CII\ dominates the cooling. As a result, the \CII/FIR ratio decreases.

We have calculated the 60/100 ratio in the region of the \CII\ map. The values range from 0.2 at the \CII\  minima to 0.5 at the \CII\ (or FIR) peaks (which trace star formation).
%These values are relatively low in comparison with those obtained by \cite{Malhotra01} and \cite{Contursi02} and reflects the low average level of star formation in M31. 
The sources of \cite{Malhotra01} and \cite{Contursi02} with a 60/100 ratio in the range 0.2-0.5 have \CII/FIR ratios of 0.4-1 $\%$. In contrast, the \CII/FIR in the region studied in this paper is constant at 2 $\%$. Therefore, no correlation is present between the \CII/FIR and the 60/100 ratios and the \CII/FIR ratio is higher than in other sources with the same 60/100 ratio \citep[Fig. 4 of ][]{Contursi02}.

\section{Physical conditions } 

In this section we compare significant line-to-line and line-to-continuum ratios
with other observations and PDR model calculations in order to derive the physical
conditions of the \CII\ emitting gas. The \CII/CO(1--0) is particularly
interesting, since it varies from  about 1300 for galactic disks
\citep{Nakagawa98,Braine_n4414c}, to  about 6000 in starbursts \citep{Stacey91} and
up to 23000 for the Large Magellanic Cloud \citep{Mochizuki94}. Even in the same
galaxy, this ratio can vary in large proportions. 
In the Milky Way, \citet{Fixsen99} 
find that the \CII/CO ratio at large scale decreases from about 2000 in the 
inner galaxy to 1000 in the outer galaxy.
Averaged over the \CII\ map, the
\CII/CO(1--0) ratio in the spiral arm of M31 is 3600, ranging from about 1500 to
6000 near the bright regions of star formation.  If the \CII\ were observed at
higher resolution still, presumably yielding a brightness distribution similar to
the 24\mum\ or H$\alpha$, the \CII/CO ratio would probably rise further still as
the CO maxima are not positioned on the maxima in star formation as traced by the 
24\mum\ or H$\alpha$.

\begin{figure}[t]
\begin{center}
\includegraphics[angle=0,width=8.5cm]{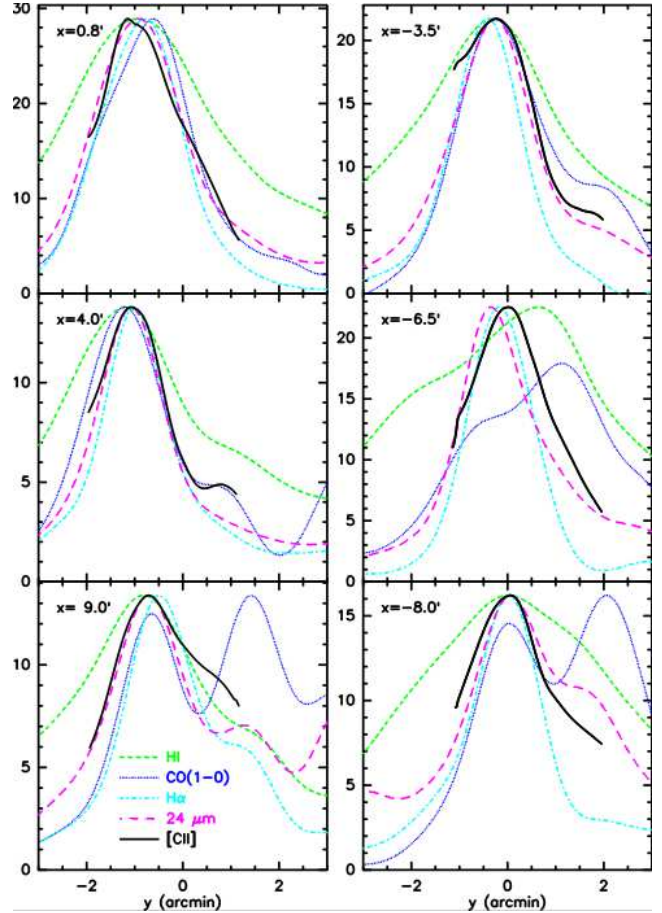}
\caption{Cuts across the spiral arm in \CII, CO(1--0), \HI, H$\alpha$, and 
the 24 \mum\ continuum.  The $x$ position refers to the horizontal axis 
position of the cut on the maps in Fig. 2. The maxima of all the cuts have been normalized to the maximum of the cut in the \CII\ image.  The units of the y-axis are those of the cut in the \CII\ image (10$^{-20}$ W\,cm$^{-2}$).  }
\label{fig_cuts}
\end{center}
\end{figure}

\begin{figure}[t]
\begin{center}
\includegraphics[angle=0,width=8.5cm]{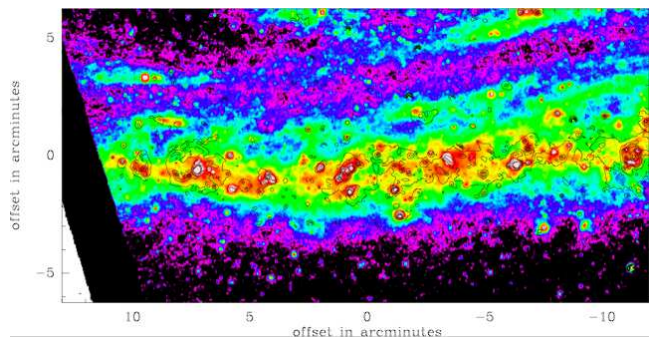}
\caption{Spitzer 24 \mum\ image with H$\alpha$ contours superposed.  Note the
excellent correspondance despite the inclination of M31 and the difference in
extinction between the two images.}
\label{fig_ha_24}
\end{center}
\end{figure}

One can use the \CII/FIR and \CII/CO ratios to study the physical conditions in the spiral arm of M31.
In addition, the \OI\ 63 \mum\ line, which comes from the dense neutral surface of PDRs, can be used in
conjunction with the \CII\ line to estimate the density of the emitting region -- the
higher the density, the higher the \OI\ to \CII\ ratio.  
We have measured \OI/\CII\  ratios of $0.51\pm0.08$ in $(x,y)\sim 
(5.5,-0.5)$ and $0.64\pm0.06$ in $(x,y)\sim (4.5,0.5)$.
Such ratios are higher than for an average spiral disk  
\citep[\OI/\CII $\sim 0.3$,][]{Braine_n4414c} and are predicted by the Kaufman et al. (1999) 
models for densities lower than the critical density of the \OI\ line 
($\sim 10^4$ \cmmt) and FUV fields at the PDR surface, $G_0$, of $30 - 300$
in units of the  FUV intensity in the local  ISM \citep[1.3 10$^{-4}$ \ui\ ,][]{Habing68} 
or for higher densities and very  low incident  fields ($G_0 \la 1$).
However, the comparison of the measured  \CII/CO(1--0) ratio with the \citet{Kaufman99} predictions constrain the PDR parameters to  $G_0=30-300$ and $n=10^3-10^{3.5}$  \cmmt. The measured (\OI+\CII)/FIR$_{42-122}$ ratio, close to 0.03, is also in agreement with the model prediction in this region of the parameters space.

However, one should take into account that a fraction of the \CII \ emission can arise from the ionized gas. In this case, one should correct the observed fluxes before comparing to the PDR models. It is difficult to estimate the amount of \CII \ that can arise from the ionized gas but statistical studies of normal galaxies \citep{Malhotra01} and the Galactic center clouds \citep{Rodriguez04} have shown that both contributions can be comparable.
Thus we have also estimated the PDR parameters assuming that the \CII\ flux that arise from PDRs is only half of the observed flux.
In this case, the comparison of the corrected \OI/\CII\ and  \CII/CO(1--0) ratios with figures 4 and 9 of \citet{Kaufman99} gives a result of 
 $G_0=30-100$ and $n=10^{3.5}-10^4$ \cmmt. The  corrected (\OI+\CII)/FIR$_{42-122}$ ratio is also consistent with this result. 
Therefore, taking into account the uncertainties on the \CII\ flux that arises in the PDR, 
the solution space from the \citet{Kaufman99} figures 4, 9, and 6 is around 
$G_0 \sim 30 - 300$ and $n$(H$) \sim 10^3 - 10^4$ cm$^{-3}$ and fits the (\CII+\OI)/FIR,
\OI/\CII, and the \CII/CO ratios.

In contrast, the predicted \CII \ intensity in this region of the 
parameter space is higher than the measured one.
For instance, the predicted intensity for $n=2000$ and $G_0=100$ 
is $\sim 2 \times 10^{-4}$ \ui, 10 times higher than that observed; the \CII\ 
intensity at the positions observed in \OI\ is 1.2-1.5 $10^{-5}$ \ui.
This can be accounted for assuming that the \CII\ emission is 
diluted in the ISO beam by a factor of $\sim 10$.
Thus, instead of $70'' \times 70''$, the angular extent of the \CII\ 
emitting regions would be $20'' \times 20''$, which is
similar to the size of the intense 
sources detected in the $5''$ (20pc) resolution 24 \mum\ image.
The \CII/FIR ratio varies very little over the region observed in \CII.
Assuming the FIR emission is also diluted by a factor 10,
one gets a corrected FIR$_{42-122}$ intensity of $\sim 10^{-2}$ \ui.  
For OB stars the FUV heating represents about half of the total grain heating
so the incident field $G_0$ is about $G_0 = 1/2 \times I_{\rm total\ FIR} / 1.3\,10^{-4}
\sim I_{\rm FIR(42-122)} / 1.3\,10^{-4} \sim 80$.
%Thus $G_0 \sim 40$.
%However, since the 42 -- 122 \mum\ FIR intensity represents only about half of 
%the total FIR intensity, the real $G_0$ deduced above is about 80.

Finally, one can estimate the column density of atomic hydrogen in the 
PDR, which can be expressed in analytical form as $N_{HI} \approx 5\,10^{20} 
\ln [90\frac{G_0}{n}+1]$ cm$^{-2}$ \citep{Sternberg89}.
For $G_0/n$ ratios of 0.01 -- 0.1, $N_{HI} \sim 3 - 12 \times 10^{20}$ \cmmd.
Roughly half of the observed \HI\ along the inner part of the spiral arm
would then be the product of photodissociation of molecular gas by the 
star formation within the arm.
PDRs observed in the Galaxy have $G_0/n \ga 0.1$, implying a greater 
\HI\ column density from photodissociation but over a much smaller area.

\begin{figure}[t!]
\begin{center}
\includegraphics[angle=0,width=8.5cm]{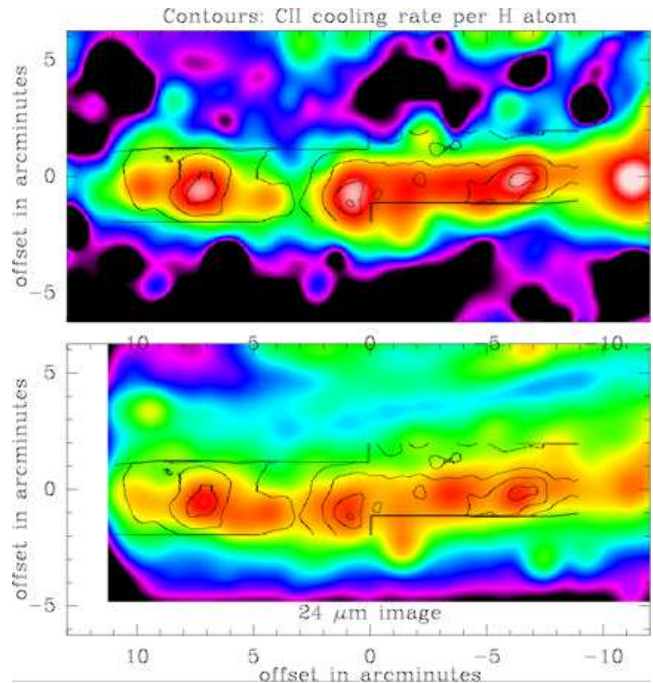}
\caption{CII cooling rate per atom of atomic Hydrogen (contours) on
the H$\alpha$ image smoothed to the \CII\ $70''$ resolution (upper panel) and
on the 24 \mum\ image smoothed to the \CII\ $70''$ resolution (lower panel).}
\end{center}
\end{figure}

\section{\CII\ cooling rate}

In order to understand how to place these observations in the general framework, we
determine the cooling rate of \CII\ per atom of \HI\ $\Lambda= 4 \pi I_{CII}/N(HI)$ and 
per H atom including molecular gas $\Lambda'= 4 \pi I_{CII}/(N(HI)+2N(H_2))$, measured at
the resolution of the \CII\ observations.  The \HI\ column density was taken from
\citet{Brinks84}, calculated using the standard formula for optically thin \HI, and we
assume a $\ratio$ factor of $2 \times 10^{20}$ cm$^{-2}$/(K\kms).  
The image of the cooling rate $\Lambda$ is plotted in Fig. 5 as contours over the
\Ha\ and 24 \mum\ smoothed images.  $\Lambda$ varies from $\sim 1.4 \,10^{-25}$
ergs\,s$^{-1}$\,atom$^{-1}$ in the maxima at $x=1.5$ and $x=-6.5$ to $\sim 2.7 \,10^{-26}$
ergs\,s$^{-1}$\,atom$^{-1}$ in the minima located around the upper border of the map.
Thus, the \CII\ emission {\it per H atom} is much higher near zones of star formation,
and varies roughly within the ranges expected for PDR gas
\citep[e.g.][]{Wolfire95,Boulanger96} but higher than expected for the CNM 
\citep{Wolfire95}.

There is little difference between $\Lambda$ and $\Lambda'$ except near the main CO
maximum because the \HI\ column dominates almost everywhere after smoothing to the 
resolution of the \CII.  Because the CO maximum is not a \CII\ maximum, the cooling rate
$\Lambda'$ is particularly low there.  The level of star formation is the important
parameter for the \CII\ emission, not the CNM column density.

\section{Conclusions}

\CII\ emission is an excellent tracer of star formation. 
The \CII\ cooling rate {\it per proton} increases greatly with the level of
star formation.  Comparison with models indicates that a coherent picture can be
obtained assuming the main source of spiral arm \CII\ emission is moderately dense
PDRs.  We estimate that the characteristic density and FUV radiation field at large 
scales are roughly 2000 cm$^{-3}$ and $G_0 \sim 100$, providing a satisfactory fit
to the \CII, \OI, CO, and FIR emission.

The \CII/FIR ratio in the spiral arm region we observed in M 31 is high,
about 2\%, well above the typically quoted values of \CII/FIR$ \approx 0.3 - 1$\% for galaxies.
This is a major result as both the \CII\ and FIR data are well-calibrated.
Looking at values and data found in the literature, the increase in
the \CII/FIR ratio with distance from the center can be seen in earlier observations
but little attention was called to the variation as the uncertainties were very
high.  The part of the arm we observed is much further out than any other \CII\
observations, which may help explain the high ratio.  Furthermore, it is clear that
in the centers of spiral galaxies the \CII/FIR ratio is lower than in the disk
so surveys of unresolved galaxies sample a combination of disk and nuclear emission
and as such can be expected to yield lower \CII/FIR values than the disks alone.

It is also
interesting that even at the $5''$ resolution of the Spitzer 24 \mum\ image, the
FIR and H$\alpha$ views of star formation are extremely similar -- very little star
formation is hidden, despite the nearly edge-on orientation of M 31. 
% Molecular
%clouds detected in CO are present with very little star formation and the \CII\
%emission is very weak there, like the 24 \mum\ and H$\alpha$, showing that massive
%stars are the important factor and not the H-atom column density.

\begin{acknowledgements} We would like to thank Nicolas Devereux for
the H$\alpha$ map of M31, Martin Haas for the ISOPHOT data, Elias Brinks 
for the H{\sc i} observations, and Nikolaus Neininger and Michel Gu\'elin
for permission to use the CO data.  
\end{acknowledgements}
\bibliographystyle{apj}
\bibliography{jb}
\end{document}